\documentclass{ws-ijmpa}

\begin{document}

\markboth{J. Vijande}{}

%
\catchline{}{}{}{}{}
%

\title{Hyperspherical harmonic formalism for tetraquarks}

\author{\footnotesize J. Vijande$^{1,2}$\footnote{E-mail address: javier.vijande@uv.es}, N. Barnea$^{3}$, A. Valcarce$^{1}$}

\address{
$^{1}$Grupo de F{\'\i}sica Nuclear and IUFFyM, Universidad de Salamanca, Salamanca, Spain\\
$^{2}$Dpto. de F\' \i sica Te\'orica and IFIC, Universidad de Valencia - CSIC, Valencia, Spain\\
$^{3}$The Racah Institute of Physics, The Hebrew University, Jerusalem, Israel\\
}

\maketitle

\begin{abstract}
We present a generalization of the hyperspherical harmonic formalism to study
systems made of quarks and antiquarks of the same flavor. This generalization
is based on the symmetrization of the $N-$body wave function
with respect to the symmetric group using the Barnea and Novoselsky algorithm.
Our analysis shows that four-quark systems with non-exotic
$2^{++}$ quantum numbers may be bound independently of the quark mass.
$0^{+-}$ and $1^{+-}$ states become attractive only for larger quarks masses.
\end{abstract}

\vspace{0.2cm}

The understanding of few-body systems relies in our capability to design
methods for finding an exact or approximate solution of the $N-$body problem.
In two-, three-, and four-body problems it is possible to obtain mathematically
correct and computationally tractable equations such as the Schr\"odinger, Faddeev and 
Yakubovsky equations describing exactly, for any assumed
interaction between the particles, the motion of few-body systems\cite{Bel90}.
However, the exact solution of these equations requires sophisticated
techniques whose difficulty increases with the number of particles.

The solution of any few-particle system may be found in a simple and unified
approach by means of an expansion of the trial wave function in terms of hyperspherical harmonic (HH)
functions. The idea is to generalize the simplicity of the spherical harmonic
expansion for the angular functions of a single particle motion to a system of
particles by introducing a global length $\rho$, called the hyperradius, and a
set of angles, $\Omega$. For the HH expansion method to be practical, the
evaluation of the potential energy matrix elements must be feasible. The main
difficulty of this method is to construct HH functions of proper symmetry for a
system of identical particles. This may be
overcome by means of the HH formalism based on the symmetrization of the
$N-$body wave function with respect to the symmetric group using the Barnea and 
Novoselsky algorithm\cite{Nir9798}. 
Therefore, this method becomes ideally suited to look for the possible existence of bound multiquark systems.

The recent series of discoveries of new meson and baryon resonances 
whose properties do not fit into the predictions of the
naive quark model, has reopened the interest on the possible role
played by non-$q\bar q$ configurations in the hadron 
spectra. Among them, the existence of four- (two quarks and two antiquarks) and five-quark 
(four quarks and one antiquark), have been suggested in the low-energy hadron 
spectroscopy. Four-quark bound states were already suggested 
theoretically thirty years ago, both in the light-quark
sector by Jaffe\cite{Jaf77} and in the heavy-quark sector by Iwasaki\cite{Iwa76}. 
In the heavy-light sector there seems to be a consensus 
that four-quark systems containing two-light and two-heavy quarks, $qq \overline Q \overline Q$, 
should be stable against dissociation into two mesons, $q\overline Q$,
if the ratio of the mass of the heavy to the light quark
is large enough\cite{Ade82}. 
Unfortunately, such an agreement do not exist for multiquarks made of four heavy quarks. 
These states have not received as much attention in the last years as their light counterparts and
the scarce theoretical predictions for the
existence of $cc\bar c\bar c$ systems differ depending basically on the method used to solve
the four-body problem and the interaction employed\cite{Iwa76,Ade82,Sil93,Llo04}.
It is our aim to make a general study of four-quark systems of identical flavor
in an exact way. For this purpose we have generalized the HH method\cite{Bar00}, 
widely used in traditional nuclear physics for the study of
few-body nuclei, to study four-quark systems. This presents two main difficulties,
first the simultaneous treatment of particles and antiparticles, and second
the additional color degree of freedom. A comprehensive discussion about the generalization 
of the HH method and the constituent quark model used can be found in Ref.\cite{Bar06}.

\begin{table}
\tbl{Charmonium spectrum in MeV. Experimental data are taken
from PDG\protect\cite{Pdg05}.}{
\begin{tabular}{|cccc|}
\hline
$(nL) \, J^{PC}$        & State                 & CQM   & Exp. \\
\hline
$(1S) \, 0^{-+}$        & $\eta_c(1S)$          & 2990  &2979.6$\pm1.2$ \\
$(1S) \, 1^{--}$        & $J/\psi(1S)$          & 3097  &3096.916$\pm0.011$ \\
$(1P) \, 0^{++}$        & $\chi_{c0}(1P)$       & 3443  &3415.19$\pm0.34$ \\
$(1P) \, 1^{++}$        & $\chi_{c1}(1P)$       & 3496  &3510.59$\pm0.10$ \\
$(1P) \, 2^{++}$        & $\chi_{c2}(1P)$       & 3525  &3556.26$\pm0.11$ \\
$(1P) \, 1^{+-}$        & $h_c(1P)$             & 3507  &3526.21$\pm$0.25 \\
$(2S) \, 0^{-+}$        & $\eta_c(2S)$          & 3627  &3654$\pm$10 \\
$(2S) \, 1^{--}$        & $\psi(2S)$            & 3685  &3686.093$\pm0.034$ \\
$(1D) \, 1^{--}$        & $\psi(3770)$          & 3776  &3770$\pm2.4$ \\ 
$(1D) \, 2^{--}$        & $\psi(3836)$          & 3790  &3836$\pm$13 \\ 
$(3S) \, 1^{--}$        & $\psi(4040)$          & 4050  &4040$\pm$10 \\ 
$(2D) \, 1^{--}$        & $\psi(4160)$          & 4104  &4159$\pm$20 \\
\hline
\end{tabular}
\label{t1}}
\end{table}  

The results for the $cc\bar c\bar c$ states have been obtained
in the framework of the hyperspherical harmonic formalism 
up to the maximum value of $K$ within our computational
capabilities ($K_{max}$). Since the only
relevant two-meson decay thresholds for these systems are those made of
two $c\bar c$ mesons, we summarize in Table \ref{t1} the results obtained for
the charmonium spectrum compared with the experimental data 
quoted by the Particle Data Group (PDG)\cite{Pdg05}. 
To analyze the stability of these systems against dissociation through strong decay,
parity ($P$), $C-$parity
($C$), and total angular momentum ($J$) must be conserved. 
In Table \ref{t3} we indicate the lowest two-meson threshold for each set of
quantum numbers. Four-quark states will be stable under strong 
interaction if their total energy lies below all possible, and allowed, two-meson 
thresholds. It is useful to define $\Delta=M(q_1q_2\bar q_3\bar q_4)-T(M_1,M_2)$
in such a way that if $\Delta>0$ the four-quark system will fall
apart into two mesons, while $\Delta<0$ will indicate that such strong decay is
forbidden and therefore the decay, if
allowed, must be weak or electromagnetic, being its width much narrower.

\begin{table}
\tbl{$cc\bar c\bar c$ masses obtained for the maximum value of $K$ computed, $E(K_{max})$, and using 
the extrapolation, $E(K=\infty)$, compared with the corresponding threshold for each set of quantum 
numbers, $M_1M_2$ and $T(M_1,M_2)$. The subindex stands for relative angular momentum in the final state. 
The value of $\Delta$ for each state is also given. All energies are in MeV.}{
\begin{tabular}{|c|cc|cc|c|}
\hline
$J^{PC}$  &$E(K_{max})$	& $E(K=\infty)$	& $M_1M_2$				&  $T(M_1,M_2)$ & $\Delta$\\
\hline
$ 0^{++}$ &     6115	&  6038  	& $\eta_c(1S)\,\,\eta_c(1S)\vert_S$	&  5980  	& +58  \\ 
$ 0^{+-}$ &     6606    &  6515         & $\eta_c(1S)\,\,h_c(1P)\vert_P$	&  6497         & +18 \\
$ 1^{++}$ &     6609    &  6530         & $\eta_c(1S)\,\,\chi_{c0}(1P)\vert_P$  &  6433         & +97 \\
$ 1^{+-}$ &     6176    &  6101         & $J/\psi(1S)\,\,\eta_c(1S)\vert_S$	&  6087         & +14 \\
$ 2^{++}$ &     6216    &  6172         & $J/\psi(1S)\,\,J/\psi(1S)\vert_S$	&  6194         & $-$22 \\
$ 2^{+-}$ &     6648    &  6586         & $\eta_c(1S)\,\,h_c(1P)\vert_P$        &  6497         & +89 \\
$ 0^{-+}$ &     7051    &  6993         & $J/\psi(1S)\,\,J/\psi(1S)\vert_P$     &  6194         & +779 \\
$ 0^{--}$ &     7362    &  7276         & $J/\psi(1S)\,\,\eta_c(1S)\vert_S$     &  6087         & +1189 \\
$ 1^{-+}$ &     7363    &  7275         & $J/\psi(1S)\,\,J/\psi(1S)\vert_P$     &  6194         & +1081 \\
$ 1^{--}$ &     7052    &  6998         & $J/\psi(1S)\,\,\eta_c(1S)\vert_S$     &  6087         & +911 \\
$ 2^{-+}$ &     7055    &  7002         & $J/\psi(1S)\,\,J/\psi(1S)\vert_P$     &  6194         & +808 \\
$ 2^{--}$ &     7357    &  7278         & $J/\psi(1S)\,\,\eta_c(1S)\vert_S$     &  6087         & +1191 \\
\hline
\end{tabular}
\label{t3}}
\end{table}

We show in Table \ref{t3} all possible $J^{PC}$ quantum numbers with $L=0$. 
Let us first of all note that the convergence of the expansion in terms of hyperspherical
harmonics is slow, and the effective potential techniques\cite{Bar00} are unable to improve it. In order 
to obtain a more adequate value for the energy we have extrapolated it according 
to the expression $ E(K)=E(K=\infty)+a/K^b$, where $E(K=\infty)$, $a$ and $b$ are fitted parameters. 
The values obtained for $E(K=\infty)$ are stable within $\pm$10 MeV for each quantum number.
A first glance to these results indicates that only three sets of quantum numbers have some probability of being observed.
These are the $J^{PC}=0^{+-}$, $1^{+-}$, and $J^{PC}=2^{++}$, which are very close to the corresponding threshold.
Being all possible strong decays forbidden these states should be narrow with typical widths of 
the order of a few MeV. It is interesting to observe that the quantum numbers $0^{+-}$
correspond to an exotic state, those whose quantum numbers cannot be obtained from a
$q\bar q$ configuration. Therefore, if experimentally observed, it would be easily 
distinguished as a clear signal of a pure non$-q\bar q$ state.

To analyze whether the existence of bound states with exotic quantum numbers could be a characteristic 
feature of the heavy quark sector or it is also present in the light sector we have calculated the value of 
$\Delta$ for different quark masses.
We have obtained that only one of the non-exotic states, the $2^{++}$, becomes more bound when the quark mass is
decreased, $\Delta\approx-80$ MeV for the light quark mass. The $1^{+-}$ and $0^{++}$
states, that were slightly above threshold in the charm sector, increase
their attraction when the quark mass is increased and only for masses close to the bottom quark mass may be bound. 
With respect to the
exotic quantum numbers, the negative parity $0^{--}$ and
$1^{-+}$ are not bound for any value of the quark mass. 
Only the $0^{+-}$ four-quark state becomes more deeply bound when the constituent quark mass increases,
and therefore only one possible narrow state with exotic quantum numbers may appear in the heavy-quark sector. This 
state present an open $P-$wave threshold only for quark masses below 3 GeV.
Let us note that since the heavy quarks are isoscalar states, the flavor wave function of the four heavy-quark
states will be completely symmetric with total isospin equal to zero. Therefore, one should 
compare the results obtained in the light-quark case with a completely symmetric flavor 
wave function, i.e., the isotensor states.

There are experimental evidences for three states with exotic quantum numbers 
in the light-quark sector. Two of them are isovectors with quantum numbers $J^{PC}=1^{-+}$ named
$\pi_1(1400)$ and $\pi_1(1600)$, 
and one isotensor $J^{PC}=2^{++}$, the $X(1600)$.
Taking the experimental mass for the threshold $T(M_1,M_2)$ in the light-quark case together with the values obtained for $\Delta$,
one can estimate the energy of these states, being $M(2^{++})\approx 1500$ MeV and $M(1^{-+})\approx 2900$ MeV.
The large mass obtained for the $1^{-+}$ four-quark state makes doubtful 
the identification of the $\pi_1(1400)$ or the $\pi_1(1600)$ with a pure multiquark state, although
a complete calculation is needed before
drawing any definitive conclusion\cite{Fut05}. 
Concerning the $X(1600)$, being its experimental mass 1600$\pm$100 MeV, a possible tetraquark configuration
seems likely.

The program we have started for an exact study of multiquark systems by means
of the HH formalism will be accomplished by implementing the possibility of
treating quarks of different masses. When this is done we will have at our
disposal a powerful method, imported from the nuclear physics, to study in an
exact way systems made of any number of quarks and antiquarks coupled to a
color singlet.

This work has been partially funded by Ministerio de Ciencia y 
Tecnolog\'{\i}a under Contract No. FPA2004-05616 and by Junta de Castilla y Le\'{o}n under
Contract No. SA-104/04.
The authors would like to thank J.-M. Richard for calling our attention about the possible existence of an $L-$wave threshold lower than $S-$wave ones.

\end{document}